\documentclass[final]{raa}            

\usepackage{graphicx,times}             
\usepackage{natbib}
\usepackage{amssymb,amsmath}
\bibpunct{(}{)}{;}{a}{}{,}

\newcommand{\km}{{~\rm km}}
\newcommand{\s}{{~\rm s}}

\newcommand{\yr}{{~\rm yr}}

\newcommand{\kpc}{{~\rm kpc}}

\newcommand{\keV}{{~\rm keV}}

\begin{document}

   \title{Imprints of the jittering jets explosion mechanism in the morphology of the supernova remnant SNR~0540-69.3
}

   \volnopage{Vol.0 (20xx) No.0, 000--000}      
   \setcounter{page}{1}          

   \author{Noam Soker
      \inst{1}
   }

   \institute{Department of Physics, Technion, Haifa, 3200003, Israel;  soker@physics.technion.ac.il {\it soker@physics.technion.ac.il}\\
\vs\no
   {\small Received~~20xx month day; accepted~~20xx~~month day}}

\abstract{
I identify a point-symmetric structure in recently published VLT/MUSE velocity maps of different elements in a plane along the line of sight at the center of the supernova remnant SNR~0540-69.3, and argue that jittering jets that exploded this core collapse supernova shaped this point-symmetric structure. The four pairs of two opposite clumps that compose this point symmetric structure suggest that two to four pairs of jittering jets shaped the inner ejecta in this plane. In addition, intensity images of several spectral lines reveal a faint strip (the main jet-axis)  that is part of this plane of jittering jets and its similarity to morphological features in a few other SNRs and in some planetary nebulae further suggests shaping by jets. My interpretation implies that in addition to instabilities, jets also mix elements in the ejecta of core collapse supernovae. Based on the point-symmetric structure and under the assumption that jittering jets exploded this supernova, I estimate the component of the neutron star natal kick velocity on the plane of the sky to be $\simeq 235 \km\s^{-1}$, and at an angle of $\simeq 47^\circ$ to the direction of the main jet-axis. I analyse this natal kick direction together with other 12 SNRs in the frame of the jittering jets explosion mechanism. 
}

 \authorrunning{N. Soker}            
   \titlerunning{The CSM of type II ILOTs}  
   
      \maketitle
\section{INTROCUTION}
\label{sec:intro}

Core collapse supernova (CCSN) remnants (SNRs) have inhomogeneous structures of filaments, arcs, clumps, and `Ears' (two opposite protrusions from the main SNR). Examples include the SNRs Vela (images by, e.g.,  \citealt{Aschenbachetal1995, Garciaetal2017}),  SNR~G$292.0+1.8$ (e.g., \citealt{Parketal2002, Parketal2007}), and SNR~W49B (e.g., \citealt{Lopezetal2013, Sanoetal2021}). 
This holds for the inner ejecta of CCSNe that have inhomogeneous structures of filaments, clumps, and rings of the different heavy elements such as oxygen, silicon, sulfur, argon and iron. 
According to the delayed neutrino explosion mechanism of CCSNe instabilities that are inherently exist in this explosion mechanism cause these filamentary structures of the inner ejecta (e.g., \citealt{Jankaetal2017, Wongwathanarat2017, Gableretal2021, Sandovaletal2021}). According to the jittering jets explosion mechanism of CCSNe both instabilities and jittering jets shape the ejecta (e.g.,  \citealt{PapishSoker2014a, PapishGilkisSoker2015, GilkisSoker2016}). In the jittering jets explosion mechanism instabilities inherently exist, in particular the spiral standing accretion shock instability (for studies of this instability see, e.g., \citealt{BlondinMezzacappa2007, Rantsiouetal2011, Fernandez2015, Kazeronietal2017}), because these instabilities supply the stochastic angular momentum to form the intermittent accretion disk that launches the jittering jets (e.g., \citealt{PapishGilkisSoker2015, ShishkinSoker2021}).

In some cases instabilities alone cannot account for the filamentary structure of the ejecta and it seems that jets as well shape the ejecta. Consider the clumpy/filamentary structure of the ejecta of SN~1987A (e.g., \citealt{Franssonetal2015, Franssonetal2016, Larssonetal2016, Abellanetal2017, Matsuuraetal2017}), now SNR~1987A. 
Although there are claims that the non-symmetric explosion of SN~1987A is due to instabilities alone (e.g.,  \citealt{Kjaeretal2010}), recent studies suggest that this is not the case. 
\cite{Abellanetal2017} find that none of the neutrino driven explosion models they compare to fit all observations of SN~1987A, a conclusion that supports a similar earlier claim by \cite{Soker2017a}. \cite{Soker2017a} compared the Fe structure of SN~1987A from \cite{Larssonetal2016} with the numerical simulations by  \cite{Wongwathanaratetal2015} and concluded that the neutrino-driven explosion mechanism cannot account for { the structure of the Fe/Si-bright regions}  of SN~1987A {(but this is in dispute in the literature, e.g., \citealt{Jankaetal2017}).} { Namely, the numerical simulations of  \cite{Wongwathanaratetal2015} predict several narrow Fe-rich fingers while the observed Fe/Si-bright structure is of two large regions. }
\cite{BearSoker2018SN1987A} use the observations of \cite{Abellanetal2017} to compare some morphological features of SN~1987A with morphological features of other SNRs and with planetary nebulae, and further argue that jittering jets played a crucial role in the explosion of SN~1987A.  
I note that the earlier claim of \cite{Wangetal2002} that two opposite non-jittering jets exploded SN~1987a is in conflict with the structure of the ejecta that \cite{Abellanetal2017} reveal (see discussion by \citealt{BearSoker2018SN1987A}). 

\cite{Wongwathanaratetal2015} and \cite{Orlandoetal2021} compared numerical simulations of the neutrino-driven explosion mechanism with the structure of SNR~Cassiopeia~A and argued that these simulations reproduce the morphological distribution of some metals in SNR~Cassiopeia~A. On the other hand, in \cite{Soker2017a} I examined the structure of Cassiopeia~A from the observations of \cite{Grefenstetteetal2017} and \cite{Leeetal2017} and showed that the metal distributions that \cite{Wongwathanaratetal2015} obtain from their numerical simulations cannot account for the observations. I instead argued that jets seem to have played a crucial role is the shaping of SNR~Cassiopeia~A during its explosion. I note also that \cite{Orlandoetal2016} argued for the existence of a large-scale asymmetrical outflow in Cassiopeia~A as instabilities alone cannot account for its morphology.

Another strong indication from SNRs to the role of jets in CCSN explosions is the presence of ears \citep{BearSoker2017b, Bearetal2017, GrichenerSoker2017}. While the distributions of heavy metals reveal mainly the SNR inner structures, ears reveal the role of jets in the outskirts of the ejecta. The reason is that most likely the ears are shaped by the last jets that the newly born neutron star (NS) has launched. This launching episode takes place after the core has exploded and therefore the jets can propagate to large distances (e.g., \citealt{Bearetal2017}). 

In the present study I examine the newly published high-quality observations and analysis of SNR~0540-69.3 by \cite{Larssonetal2021} to argue that there is a point-symmetric structure as expected in the jittering jets explosion mechanism (section \ref{sec:imprints}).
X-ray observations (e.g. \citealt{Parketal2010}) show that SNR~0540-69.3 has a large-scale rectangle shape, but one that is highly non-homogeneous. The $2.5-7 \keV$ intensity map, however, contains two bright spots on opposite sides of the center (one to the east and one to the west; figure 1 of \citealt{Parketal2010}). This pair might be or not part of the point symmetric structure that I study here. However, because the two bright spots are on the edge of the SNR and are faint, I do not explore their nature here. 

In section \ref{sec:kick} I use my claim for jittering jets in a plane to estimate the projected angle between the pulsar natal kick direction and the main jet-axis of the jittering jets. I incorporate this angle with 12 other SNRs for which the projected angles of the jets to the kick velocity exist. I use these 13 angles to strengthen an earlier claim that the natal kick velocity avoids small angle with respect to the main jet-axis. 
I discuss and summarise my study in section \ref{sec:summary}.

\section{Imprints of jittering jets}
\label{sec:imprints}
\subsection{The main jet-axis}
\label{subsec:axis}
   
\begin{figure*}[ht!]
\includegraphics[trim=0.6cm 17.2cm 0.0cm 1.2cm ,clip, scale=0.80]{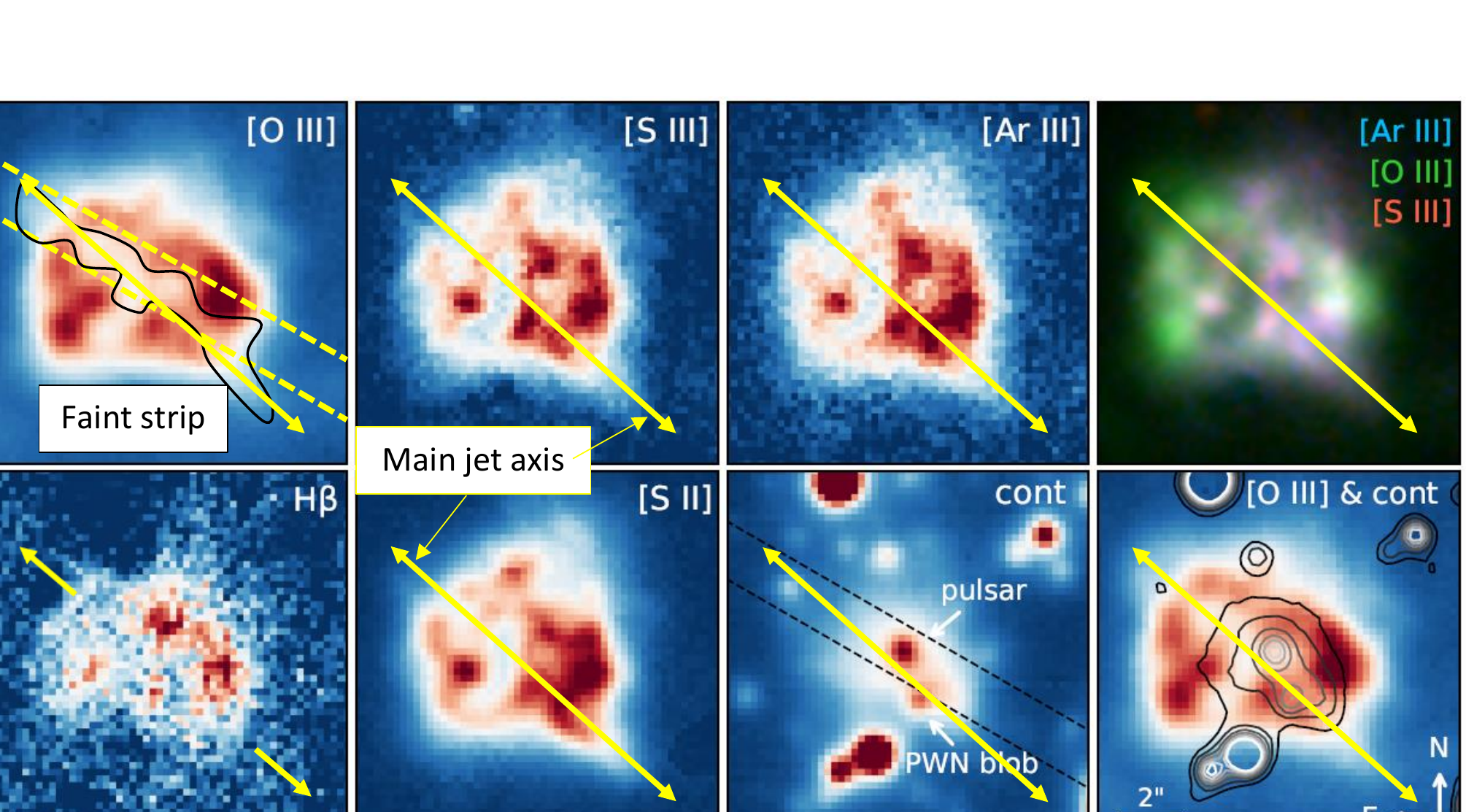} \\
\caption{My claim for a main jet-axis on the plane of the sky as I mark with yellow double-headed arrows on the images reproduced from figure 1 of \cite{Larssonetal2021}. My additions are the yellow arrows, the dashed lines on the upper left panel that are the same as the dashed black line on the third lower panel, and the mark of the faint-strip by a black line in the upper left panel.
The two parallel dashed lines mark the boundary of the $1.5^{\prime \prime}$ wide X-shooter slit (\citealt{Larssonetal2021}). }
	\label{fig:LarssonFig1}
\end{figure*}

The high resolution { VLT/MUSE } observations and high quality three-dimensional  { reconstruction of the ejecta } of SNR~0540-69.3 by \cite{Larssonetal2021} allow the identification of a point-symmetric morphology. I first present in Fig. \ref{fig:LarssonFig1} the images as \cite{Larssonetal2021} do in their figure 1. On these images I added yellow double-headed arrows along a faint strip that goes through the center of the SNR as I mark by the black solid line in the upper left panel of Fig. \ref{fig:LarssonFig1} (all yellow arrows on the different panels are on the same position). I term this axis the \textit{main jet-axis}, but as I discuss below in might represent a plane along the line of sight. From comparison of such central faint strips in planetary nebulae and in SNRs earlier studies (e.g., \citealt{Bearetal2017, Akashietal2018}) suggested that the faint strip in SNRs is the direction of two opposite jets that cleaned the axis from most of the gas. Here, as well, I suggest hat this central faint strip marks the direction of two opposite jets, or even two to four such pairs of jets in a plane. Namely, as I show in section \ref{subsec:pointsymmetry} more likely there is a jittering plane that represents two or more pairs of jittering jets. I also note that the double-headed arrows that mark the faint strip do not go through the pulsar, but rather through the pulsar wind nebula (PWN) blob. However, the pulsar is inside the faint strip as the strip is wide, but not at its central axis. 

\begin{figure*}[ht!]
\includegraphics[trim=0.0cm 16.1cm 0.0cm 1.2cm ,clip, scale=0.75]{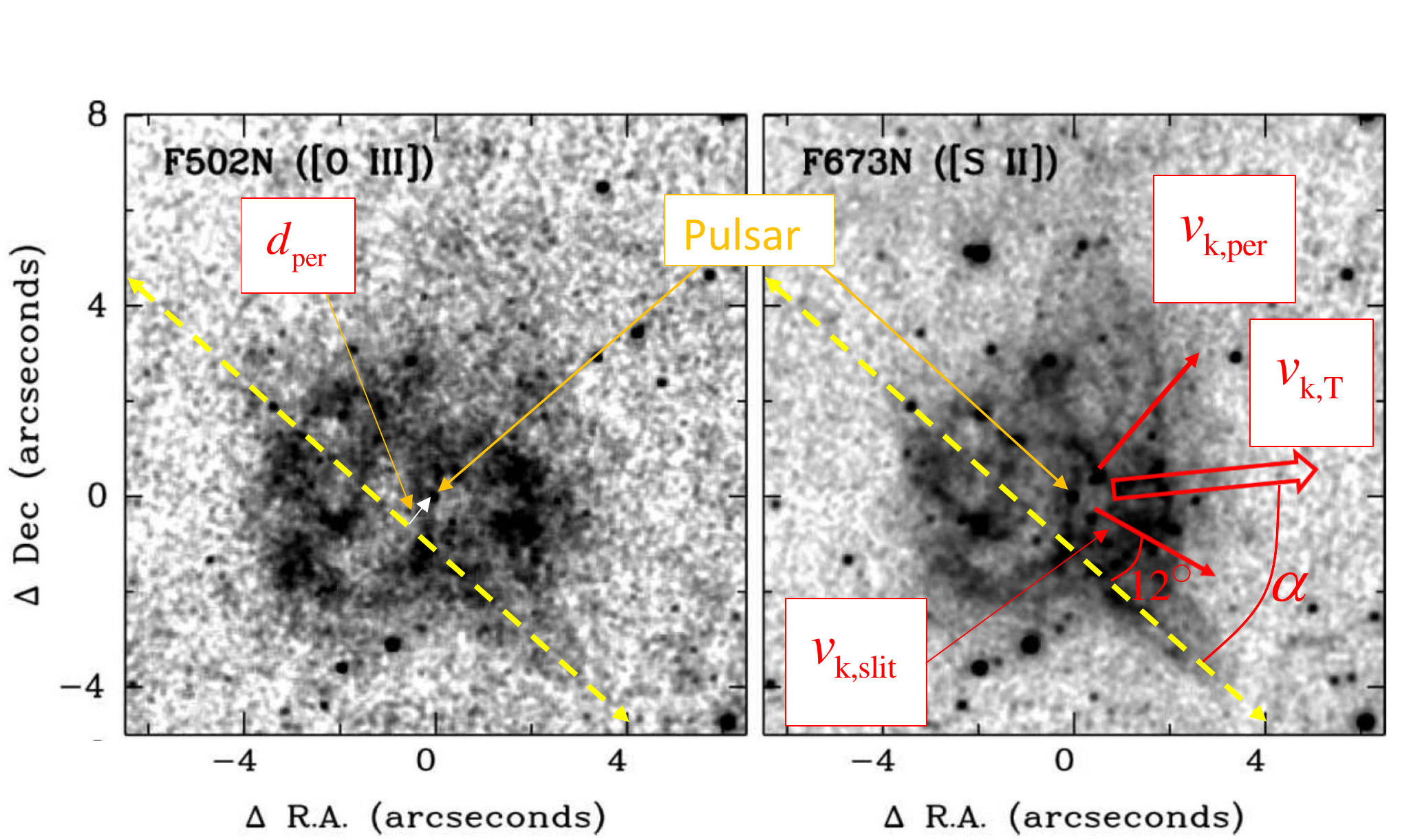} \\
\caption{My identification of the main jet-axis (yellow double-headed arrows) in the HST observations from \cite{Morseetal2006}. In red double-lined arrow I mark my suggestion for the transverse (projected on the plane of the sky) NS natal kick velocity $v_{\rm k,T}$ that I derive from the velocity component perpendicular to the main jet-axis $v_{\rm k,per}$ and the component along the slit $v_{\rm k,slit}$ (section \ref{sec:kick}).  }
	\label{fig:MorseFig1}
\end{figure*}

I can identify the same main jet-axis in earlier Hubble Space Telescope (HST) observations by \cite{Morseetal2006}, as I mark on the two panels of Fig. \ref{fig:MorseFig1} that I reproduce from figure 1 of \cite{Morseetal2006}. I discuss the meaning of the red arrows of the NS natal kick velocity in section \ref{sec:kick}. 

My identification of the faint strip as the axis (or plane) of jittering jets that exploded the CCSN is new. In the past, studies (e.g., \citealt{Serafimovichetal2004, Brantsegetal2014}) attributed this plane to the plane of the assumed torus of the PWN.  These studies also identified two opposite jets perpendicular to this torus (e.g., \citealt{Serafimovichetal2004, Lundqvistetal2021}). On the other hand, \cite{DeLucaetal2007} suggested that the direction along the faint strip is the direction of the pulsar jets (see also discussion by \citealt{Lundqvistetal2011}). As I show next, I attribute this direction to the plane of the jittering jets that exploded the CCSN of  SNR~0540-69.3 rather than to the jets of the pulsar. 

\subsection{A point-symmetric Morphology}
\label{subsec:pointsymmetry}

\subsubsection{Point-symmetry in planetary nebulae}
\label{subsubsec:PNe}

\cite{Larssonetal2021} present a thorough analysis of the properties of the clumps and the rings that they reveal in SNR~0540-69.3. I here only concentrate on what I identify as a point-symmetric structure that I attribute to jittering jets.
My claim for jets is based in large parts on the shaping by jets of the point-symmetric morphologies of planetary nebulae (e.g. \citealt{SahaiTrauger1998, Sahaietal2011}).  

\begin{figure*}[ht!]
\includegraphics[trim=0.0cm 16.2cm 0.0cm 2.0cm ,clip, scale=0.75]{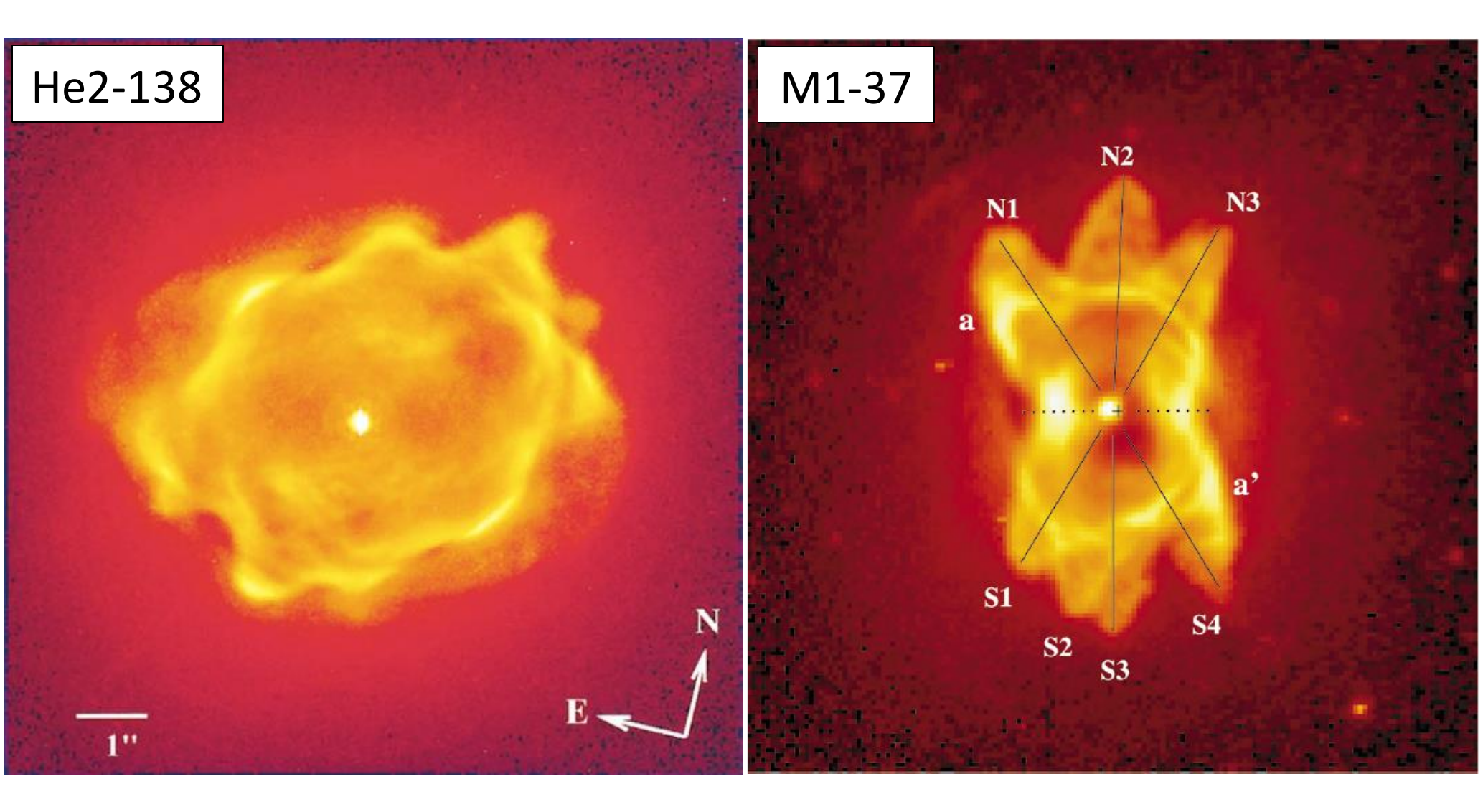} \\
\caption{ {  Two-planetary nebulae that demonstrate a point-symmetric structure that the authors of the respective papers attribute to jets. On the left is the PN He2-138 (PN~G320.1-09.6) from \cite{SahaiTrauger1998} and on the right is PN~M1-37 (PN~G002.6-03.4) from \cite{Sahai2000}. The three lines on the right panel are on the original image by \cite{Sahai2000}. }
}
\label{fig:PNe}
\end{figure*}

{ In Fig. \ref{fig:PNe} I present two examples of planetary nebulae with a point-symmetric structure. On the left is the planetary nebula He2-138 (PN~G320.1-09.6) from \cite{SahaiTrauger1998} that presents a point symmetric structure that the authors attribute to shaping by jets. This PN demonstrates four pairs of opposite protrusions. In this case most of the brightest regions are the arcs at the front of the protrusions. On the right is the planetary nebula PN~M1-37 (PN~G002.6-03.4) from \cite{Sahai2000}, who also marked the three straight lines and named the three pairs of protrusions (lobes). In this case most of the brightest regions are to the sides of the lines connecting the tips of the opposite protrusions.  
}  

{ The point to take from these two planetary nebulae is that jets can form point-symmetric structures, but that the brightest regions might be on different regions with respect to what researchers identify as the jets' axes.  }

\subsubsection{Point-symmetry in SNR~0540-69.3}
\label{subsubsec:psSNR}

\begin{figure*}[ht!]
\includegraphics[trim=0.2cm 13.2cm 0.0cm 1.5cm ,clip, scale=0.75]{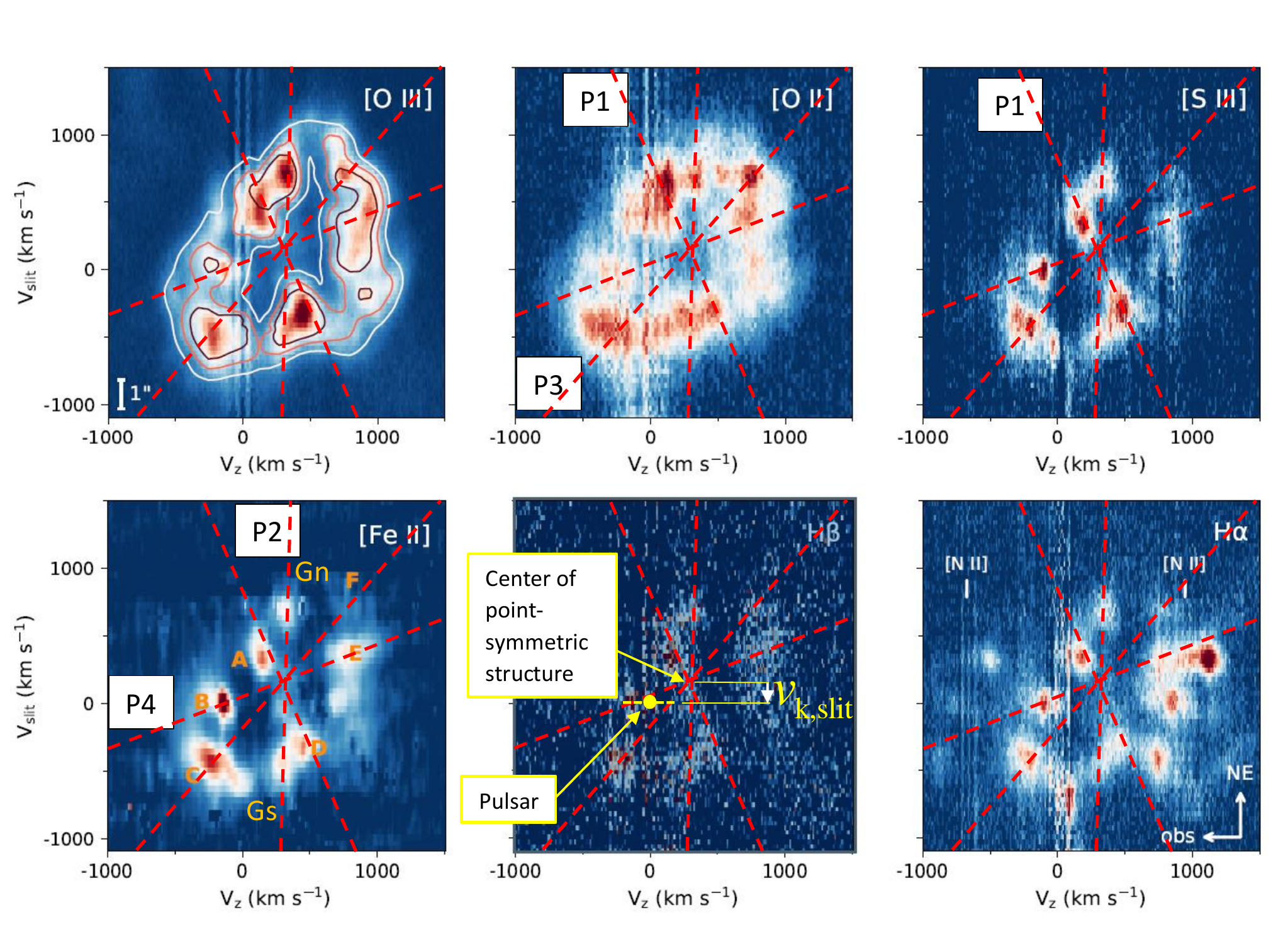} \\
\caption{Two-dimensional velocity maps as \cite{Larssonetal2021} present in their figure 4 along the slit that the two dashed lines mark in two panels of Fig. \ref{fig:LarssonFig1}.
The velocity along the slit $v_{\rm slit}$ is positive to the northeast, and $v_{\rm z}$ is the velocity along the line of sight. 
I added 4 lines on all panels (same 4 lines), P1-P4, to mark 4 pairs of opposite clumps that form the point-symmetric morphology that I identify here (A-D; B-E; C-F; Gn-Gs). The marks of clumps A to F in the lower left panel are from \cite{Larssonetal2021}, while I added the marks to clumps Gn and Gs. The pulsar is at $v_{\rm slit}=0$ in these panels. { The dashed horizontal yellow line at the location of the pulsar indicates that the line of sight velocity of the pulsar is not determined here.} }
	\label{fig:PointSymmetry}
\end{figure*}

From their VLT/X-shooter spectroscopic observation along the slit as marked by two parallel dashed lines in two panels of Fig. \ref{fig:LarssonFig1} and from their derived age of $\simeq 1100 \yr$ for SNR~0540-69.3, \cite{Larssonetal2021} build a velocity map in a plane perpendicular to the plane of the sky, i.e., along the line of sight. I present their figure 4 in Fig. \ref{fig:PointSymmetry}. I identify 8 clumps that form a point-symmetric structure (although not perfect) along four lines that go through four pairs of clumps A-D, B-E, C-F and Gs-Gn. 
\cite{Larssonetal2021} mark clumps A-F and I mark also clumps Gn and Gs.  

{ I define the four different lines by connecting bright clumps. I identify the line `P1' in the images of [O II] and [S III], the line `P2' in the image of [Fe II], the line `P3' in the image of the [O II], and the line `P4' in the image of [Fe II]. I then copy the lines to other images to form the same structure of the 4 lines in all images of Fig. \ref{fig:PointSymmetry}. The strong point is that the lines cross also clumps that I did not use to define them. For example, line `P2' crosses also clumps in the [O II] image and in the H$\alpha$ image. }  

{ Because I define the lines by the bright clumps, the 4 lines do not cross exactly at the same point. However, that the four lines cross each other at almost the same point, despite that I define them by the clumps, supports my claim for a point-symmetric structure.  }

A point-symmetric structure of an outflowing nebula very strongly suggests shaping by precessing jets or jittering (stochastic) jets. I therefore take the point symmetric structure in the plane of the slit to be a plane of jittering jets. The slit direction is at $12^\circ$ to the main jet-axis that I take along the faint strip (Fig. \ref{fig:MorseFig1}). Because the faint strip is wide, to the accuracy of the present analysis I can take the slit direction to be as the main jet-axis. This implies that the main jet-axis represent a jittering plane along the line of sight (perpendicular to the planes of Figs. \ref{fig:LarssonFig1} and \ref{fig:MorseFig1}). 
 
{ From the two point-symmetric planetary nebulae in Fig. \ref{fig:PNe} we learn that the dense clumps might be at the tip of the jets or to to the sides o the jets. } Without numerical simulations of the explosion (which are extremely demanding) I cannot tell whether there were two, three or fours pairs of jittering jets in the plane of Fig. \ref{fig:PointSymmetry}. The reason is that each jet might form a clump at its head, a case that implies four pairs of jets, or each jet can inflate a bubble that form clumps on its boundary, a case that allows for only two pairs of jets. 

Let me consider the case where four pairs of jittering jets formed the point-symmetric structure that I identify in the plane of Fig. \ref{fig:PointSymmetry}. According to the jittering jets explosion mechanism several to few tens of jet-launching episodes explode the star (e.g. \citealt{PapishSoker2011, PapishSoker2014a}). However, the jittering direction might not be completely chaotic. \cite{PapishSoker2014bPlanar} conducted three-dimensional hydrodynamical numerical simulations of the jittering jets explosion mechanism and showed that early jittering jets channel the gas that the newly born NS continues to accrete to inflow in directions perpendicular to the early jets. The direction of the angular momentum that this in-flowing gas carries, therefore, tends to be in the same plane as the first two pairs of jittering jets. Namely, later jets are more likely to be in the same plane as the first four jets (two pairs of jets). For that, the presence of four pairs of jets in a similar plane is compatible with the jittering jets explosion mechanism. This must not be the case with all jets, as large fluctuations of the angular momentum of the accreted gas onto the NS might tilt the accretion disk by a large angle so that the jet-axis of the newly launched jets is outside the earlier jittering plane. Therefore, other jet-axes are possible beside the jittering plane of Fig. \ref{fig:PointSymmetry}. In section \ref{sec:summary} I suggest that one such jet-axis might be perpendicular to the [O III] irregular ring-like structure. 
 
My conclusion is that the point-symmetric morphology that I identify in Fig. \ref{fig:PointSymmetry} is compatible with, and strongly support, the jittering jets explosion mechanism. 

\section{On the natal kick direction}
\label{sec:kick}

In what follows I deal only with the angle between the pulsar natal kick direction and the main jet-axis as it is projected on the plane of the sky. Therefore, the natal kick velocity component along the line of sight is not relevant.

\cite{Serafimovichetal2004} estimated a pulsar transverse velocity of $1190 \pm 560 \km \s^{-1}$ in a south east direction. However,  \cite{Mignanietal2010} constraint the transverse velocity to be $<250 \km s^{-1}$.

If my identification of the jittering jet axis/plane holds, I can use it to estimate the pulsar natal kick velocity { component on the plane of the sky.} The pulsar is at a transverse distance of $d_{\rm per} \simeq 0.8^{\prime \prime} D$ from the main jet-axis. For an age of $\tau= 1100 \yr$ as \cite{Larssonetal2021} use (note that \citealt{Serafimovichetal2004} take $\tau = 1660 \yr$), and a distance to the CCSN of $D=50 \kpc$ this displacement corresponds to a transverse kick velocity component perpendicular to the main jet-axis of $v_{\rm k,per} = d_{\rm per} / \tau= 171 \km \s^{-1}$. I mark this direction by thick red arrow on Fig. \ref{fig:MorseFig1}. 
\cite{Larssonetal2021} take the pulsar at the center of the explosion. From Fig. \ref{fig:PointSymmetry} I find the velocity along the X-shooter of the pulsar relative to the center of the point-symmetric structure to be $v_{\rm k,slit} = 165 \km \s^{-1}$ in the south-west direction, as I mark by a second red arrow on Fig. \ref{fig:MorseFig1}, {and by the white arrow in the middle-lower panel of Fig. \ref{fig:PointSymmetry}.} 
The slit is tilted at $12^\circ$ to the main jet-axis. From these I find the transverse (i.e., projected on the plane of the sky) pulsar velocity relative to the center of the point-symmetric structure to be { $v_{\rm k,T}=\vert \vec {v}_{\rm k,slit} + \vec {v}_{\rm k,per} \vert=235 \km \s^{-1}$. } I mark the transverse pulsar velocity that I estimate here by a double-lined red arrow in Fig. \ref{fig:MorseFig1}. This velocity is below the upper limit that \cite{Mignanietal2010} deduce. 
  
The angle of this transverse kick velocity to the main jet-axis is $\alpha \simeq 47^\circ$. I also note that the transverse component of the natal kick velocity of the pulsar that I deduce here is almost opposite to the direction that \cite{Serafimovichetal200S} argued for and that was refuted by \cite{Mignanietal2010}.
 
\begin{figure}[!t]
\includegraphics[trim=3.6cm 8.2cm 0.0cm 8.0cm ,clip, scale=0.63]{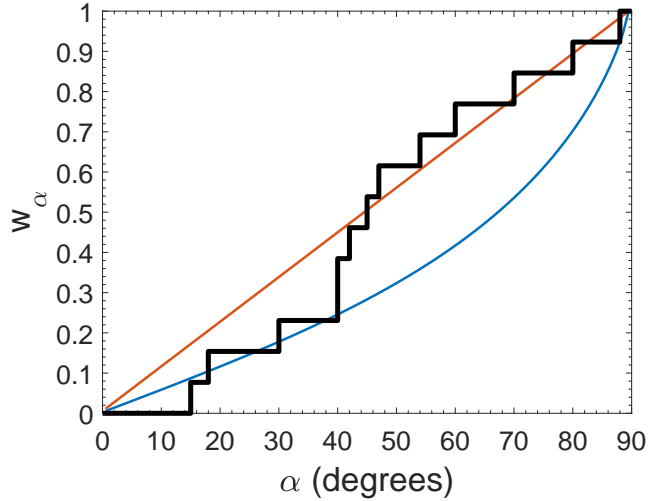}
\caption{The cumulative distribution function ${\rm W}_{\alpha}$ of projected angles between the jets'-axis and the NS natal kick for 13 SNRs. The angles of 12 SNRs are from \cite{BearSoker2018kick} and the new addition is my estimate of $\alpha \simeq 47^\circ$ for SNR~0540-69.3. The straight orange line is the expected random cumulative distribution function, while the convex blue line is the expected cumulative distribution function when in all cases the three-dimensional NS kick velocity is perpendicular to the jets' axis.
}
\label{fig:distribution}
\end{figure}

\cite{BearSoker2018kick} present the distribution of the projected angles $\alpha$ between the jets’ axis and the NS kick velocity for 12 SNRs. They conclude that the cumulative distribution function fits the random distribution (kick velocity is random with respect to the main jet-axis) at large angles, and is missing systems with small angles relative to the random distribution. 
I add the angle of $\alpha = 47^\circ$ for SNR~0540-69.3 that I estimate above to have now a sample of 13 SNRs. Note again that I deal here only with the projected angle on the plane of the sky as \cite{BearSoker2018kick} do in their analysis, and for that the natal kick velocity component along the line of sight is not relevant. I present the new cumulative distribution function in Fig. \ref{fig:distribution}.  The new addition of SNR~0540-69.3 is compatible with the conclusion of \cite{BearSoker2018kick} and strengthens it.  

The reason that in the jittering jets explosion mechanism the NS kick velocity tends to be at a large angle to the main jet-axis is that dense ejecta clumps accelerate the NS by the gravitational tug-boat mechanism (\citealt{Nordhausetal2010, Wongwathanaratetal2013, Janka2017}; for a different explanation for kick velocities see, e.g., recent studies by \citealt{Yaoetal2021} and \citealt{Xuetal2021}). \cite{BearSoker2018kick} argue that either the jets prevent the formation of dense clumps along their propagation direction, or that the dense clumps also supply the gas to the accretion disk  that launches the jets. In either cases the jet-axis and the natal kick velocity direction cannot be too close to each other.    

\section{Discussion and Summary}
\label{sec:summary}

I examined the images and velocity maps of SNR~0540-69.3 that \cite{Larssonetal2021} present in a recent study. I defined a faint strip (upper left panel of Fig. \ref{fig:LarssonFig1}) in the images of the different spectral lines, and based on similar structures in a few other SNRs and in some planetary nebulae (e.g., \citealt{Akashietal2018}) I attributed the shaping of the faint strip to one or more pairs of opposite jets. I termed this axis the main jet-axis.  

\cite{Larssonetal2021} obtain the velocity maps, which I present here in Fig. \ref{fig:PointSymmetry}, in a plane perpendicular to the plane of the sky by a slit that the two dashed lines in two panels of Fig. \ref{fig:LarssonFig1} mark. 
In the velocity maps I identified a point-symmetric structure that is defined by 8 clumps (Fig. \ref{fig:PointSymmetry}). The four lines that connect opposite clumps cross each other at what I identified as the center of the structure. 
The direction of the slit almost coincides with the main jet-axis, and therefore I take the main jet-axis to be part of the point symmetric structure. Namely, the main jet-axis was shaped by two to four pairs of jittering jets. 

In section \ref{subsec:pointsymmetry} I argued that the point symmetric structure in the plane of Fig. \ref{fig:PointSymmetry} is compatible with the jittering jets explosion mechanism of CCSNe. Actually, the jittering jets explosion mechanism predicts the common occurrence of point-symmetric morphological features in remnants of CCSNe. More than that, according to the jittering jets explosion mechanism in some cases several consecutive pairs of jets will jitter in the same plane \citep{PapishSoker2014bPlanar}.   
According to the interpretation I suggest here the elongation of the PWN (e.g., \citealt{Brantsegetal2014}) in the same direction as the main jet-axis results from the process by which the PWN plasma fills the less dense volume that the jittering jets that exploded the star shaped (the faint strip).  

From the location of the pulsar relative to the main jet axis and the age of $1100 \yr$ for SNR~0540-69.3 that \cite{Larssonetal2021} estimate, I estimated the transverse (projected on the plane of the sky) pulsar natal kick velocity to be $v_{\rm k,T} \simeq 235 \km \s^{-1}$ at $\alpha \simeq 47^\circ$ to the direction of the main jet-axis (red double-lined arrow on the right panel of Fig. \ref{fig:MorseFig1}). 

In Fig. \ref{fig:distribution} I present the cumulative distribution function of the jet-kick angles of 13 SNRs, 12 SNRs from \cite{BearSoker2018kick} and the new addition of $\alpha \simeq 47^\circ$ for SNR~0540-69.3. This distribution shows that the NS natal kick direction and the main jets'-axis avoid small angles with respect to each other. I discussed in section \ref{sec:kick} the explanation for the missing small values of $\alpha$ in the frame of the jittering jets explosion mechanism.    

Each jet inflates a bubble as it interacts with the core material that it accelerates, and the interactions with each other of the bubbles that the jittering jets inflate form a complicated flow structure in the exploding core, i.e., vortexes \citep{PapishSoker2014bPlanar}. This implies that in addition to instabilities (e.g., \citealt{Utrobinetal2019}) that occur also in the jittering jets explosion mechanism, the jittering jets also mix elements in the exploding core.   

I here analysed only the inner ejecta with expanding velocities from the explosion site of $\la 1000 \km \s^{-1}$. I did not analyse the [O III] irregular ring-like structure at a velocity of $\simeq 1600 \km \s^{-1}$ that \cite{Larssonetal2021} study in details. 
\cite{Larssonetal2021} mention that the [O III] irregular ring-like structure of SNR~0540-69.3 might be similar to the  CO torus expanding with a velocity of $\simeq 1700 \km \s^{-1}$ that ALMA observations reveal in SNR~1987A \citep{Abellanetal2017}. \cite{BearSoker2018SN1987A} attributed the shaping of the CO torus in SNR~1987A to jittering jets. I therefore propose here the possibility that the [O III] irregular ring-like structure of SNR~0540-69.3 was compressed by two opposite jets with a jets' axis perpendicular to the plane of the [O III] irregular ring-like structure. This pair of jets was launched at a different angle than the 4 pairs of jets in the plane of Fig. \ref{fig:PointSymmetry}. 

I also note that in a recent paper \cite{Leonardetal2021} interpret their polarisation observation of Type II-P/L SN 2013ej as resulting from the formation of high velocity ($\simeq 4000 \km \s^{-1}$) nickel-56 clumps in the explosion. The jittering jets might explain such fast clumps. 
  
\cite{Larssonetal2021} write in their conclusions that their results ``add to the growing evidence that rings and clumps are ubiquitous features of SN ejecta, likely reflecting hydrodynamical instabilities in the explosions.''
I would add that these clumps, rings, and mixing reflect hydrodynamical instabilities and \textbf{jets} in the explosion mechanism. 

\section*{Acknowledgments}
 
{ I thank an anonymous referee for good suggestions. } 
This research was supported by a grant from the Israel Science Foundation (769/20).



\label{lastpage}
\end{document}